%% file: main.tex
\documentclass[sigplan,10pt]{acmart}

\AtBeginDocument{%
  }

\usepackage{tikz}
\usepackage{amsmath}

\usepackage{filecontents}

\usepackage{pifont}
\usepackage{newtxtext,newtxmath}
\usepackage{url}
\usepackage{xcolor}
\usepackage{listings}
\usepackage{algorithm}
\usepackage{algorithmic}
\usepackage{paralist}
\usepackage{booktabs}

\newcommand{\rpp}{\textsc{RoundPipe}}
\newcommand{\cdp}{Computation Dispatch Paradigm}
\newcommand{\lmcp}{Layerwise Model Consistency Protocol}

\newif\ifrevision
\revisionfalse
\newcommand{\revcolor}{\ifrevision\color{blue}\fi}
\newcommand{\rev}[1]{{\revcolor #1}}
\newenvironment{revblock}{\revcolor}{}
\lstdefinestyle{base}{
moredelim=**[is][\color{blue}]{@}{@},
moredelim=**[is][\color{red}]{^}{^},
}

\copyrightyear{2026}\acmYear{2026}
\setcopyright{cc}
\setcctype{by}
\acmConference[SIGOPS ATC 2026]{The 2026 ACM SIGOPS Annual Technical Conference}{November 15-18, 2026}{Hong Kong, China}
\acmBooktitle{The 2026 ACM SIGOPS Annual Technical Conference (SIGOPS ATC 2026), November 15-18, 2026, Hong Kong, China}
\acmDOI{10.1145/3828523.3857063}
\acmISBN{979-8-4007-2818-1/2026/11}



\begin{document}

\date{}

\title{Efficient Training on Multiple Consumer GPUs with \rpp{}}

\author{Yibin Luo, Shiwei Gao, Huichuan Zheng, Youyou Lu, Jiwu Shu}
\authornote{Jiwu Shu is the corresponding author (shujw@tsinghua.edu.cn).}
\affiliation[obeypunctuation=true]{%
  \institution{Tsinghua University}\country{}
}
\renewcommand{\shortauthors}{Luo et al.}


\acmSubmissionID{769}

\begin{abstract}
Fine-tuning Large Language Models (LLMs) on consumer-grade GPUs is highly cost-effective, yet constrained by limited GPU memory and slow PCIe interconnects. Pipeline parallelism combined with CPU offloading mitigates these hardware bottlenecks by reducing communication overhead. However, existing PP schedules suffer from an inherent limitation termed the \textit{weight binding} issue. Binding uneven model stages (e.g., the LM head is large) to GPUs limits the pipeline's throughput to that of the GPU with the heaviest load, leading to severe pipeline bubbles.

In this paper, we propose \rpp{}, a novel pipeline schedule that breaks the weight binding constraint on consumer GPU servers. \rpp{} treats GPUs as a pool of stateless execution workers and dynamically dispatches computation stages across devices in a round-robin manner, achieving a near-zero-bubble pipeline. To ensure training correctness and system efficiency, \rpp{} integrates a priority-aware transfer scheduling engine, a \lmcp{}, and an automated layer partitioning algorithm. Evaluations on an 8$\times$ RTX 4090 server demonstrate that \rpp{} achieves 1.48--2.16$\times$ speedups over state-of-the-art baselines when fine-tuning 1.7B to 32B models. Remarkably, \rpp{} enables LoRA fine-tuning of the Qwen3-235B model with 31K sequence length on a single server.

\rpp{} has been open-sourced as a Python library.

\noindent \raisebox{0.1ex}{\includegraphics[height=1.5ex]{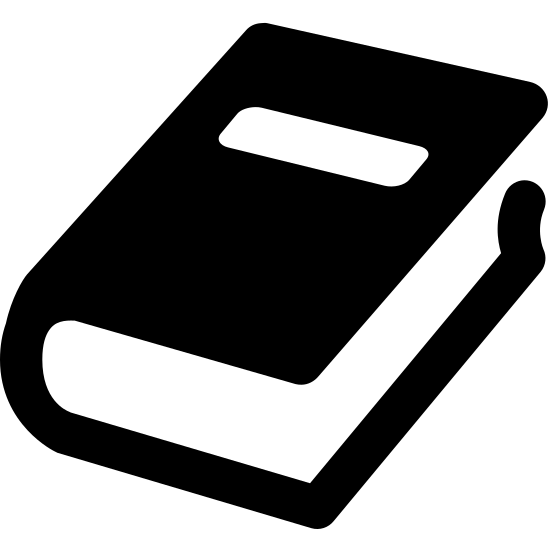}}
Project Page: \url{https://thustorage.github.io/RoundPipe/}

\end{abstract}

\maketitle
\pagestyle{empty}
\thispagestyle{empty}



\input{sec/intro}

\input{sec/background}
\input{sec/method}

\input{sec/design}
\input{sec/eval}
\input{sec/discussion}
\input{sec/related}

\section{Conclusion}
\label{sec:conclusion}

This paper presents \rpp{}, a system that implements a new pipeline parallelism schedule for training large models on consumer GPU servers.
We introduced the \cdp{} and confirmed that it preserves full compute-bound throughput.
Building on this paradigm, the \rpp{} schedule uses asymmetric stage splitting and round-robin dispatch to decouple stages from GPUs, mitigate stage imbalance, and improve pipeline efficiency.
Experiment results demonstrate its performance gain.

\section*{Acknowledgments}

This work is supported by the National Key R\&D Program
of China (Grant No. 2024YFB4505201), the National Natural
Science Foundation of China (Grant No. U22B2023,
62332011), and the High Performance Computing Center
of Tsinghua University.

\bibliographystyle{ACM-Reference-Format}
\bibliography{ref}

\end{document}

%% file: sec/intro.tex
\section{Introduction}
\label{sec:intro}

Fine-tuning Large Language Models (LLMs) has become the cornerstone of modern AI applications, enabling open-source foundation models to master complex reasoning and domain-specific tasks~\cite{ouyang2022traininglanguagemodelsfollow, roziere2023code}. Unlike pre-training from scratch, fine-tuning requires significantly less computation~\cite{devlin2019bert}. For such workloads, consumer-grade GPUs offer a compelling alternative.
The NVIDIA RTX 4090 delivers computation capabilities comparable to those of datacenter A100 GPUs, while costing roughly $80\%$ less.
Efficiently fine-tuning massive LLMs on consumer GPU servers is vital for democratizing AI for small companies and researchers.



However, consumer-grade GPUs impose two hardware constraints that restrict model scalability and training efficiency.
(1) Limited memory capacity: The VRAM of a typical consumer-grade GPU falls short of training demands. For instance, training an 8B model needs 128 GB for model states alone~\cite{rajbhandari2020zeromemoryoptimizationstraining}, far larger than 24GB for NVIDIA RTX 4090 or 32GB for NVIDIA RTX 5090. Furthermore, expanding context windows for complex tasks~\cite{Guo_2025,wan2025wanopenadvancedlargescale} (e.g., reasoning, video generation) also inflates activation memory.
(2) Slow inter-GPU communication: Consumer-grade GPUs use PCIe interconnects, offering less than 20\% of NVLink bandwidth. This physical limitation is further compounded by root complex contention in PCIe topologies~\cite{feng2023mobius,kim2024tccl}.
\begin{figure*}[t!]
\centering
\includegraphics[width=474.07px]{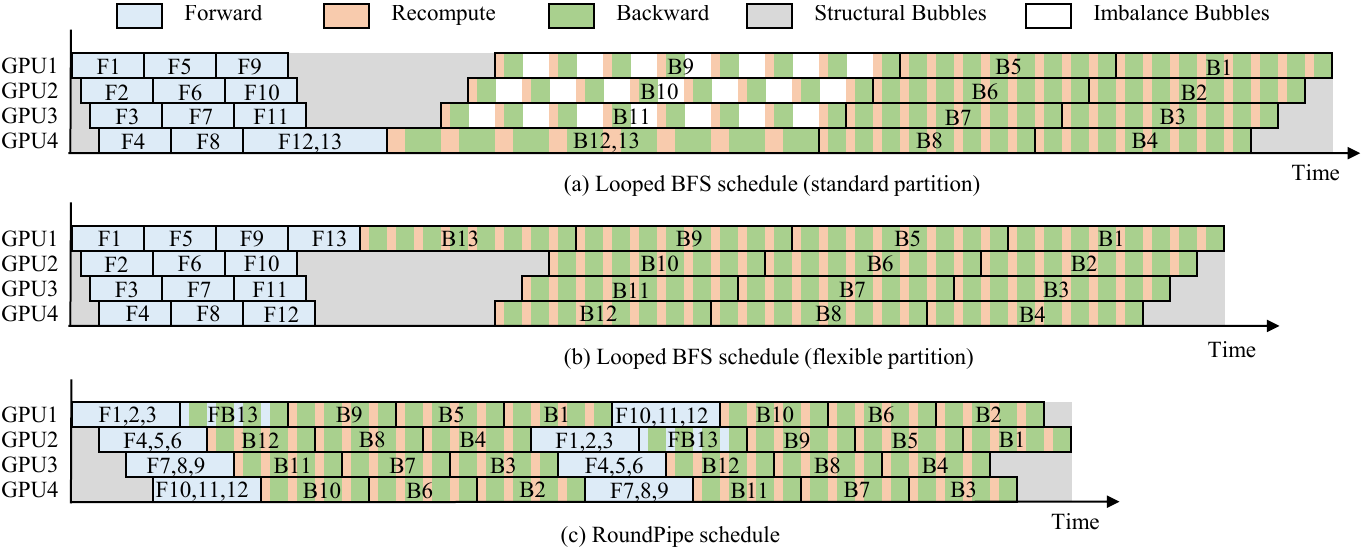}
\caption{Looped BFS schedule and \rpp{} schedule when training a 12-layer model with a language model head (layer 13) on 4 GPUs. F/B denotes forward/backward, and numbers indicate the index of layers contained in the stage. Looped BFS schedule processes 8 microbatches at a time while \rpp{} processes them in two rounds.}
\label{fig:rpp-schedule}
\end{figure*}

To alleviate VRAM constraints during consumer-grade GPU training, existing methods commonly employ CPU offloading~\cite{ren2021zerooffloaddemocratizingbillionscalemodel, rajbhandari2021zeroinfinitybreakinggpumemory} to shift parameters, optimizer states, and activations to host memory, often combined with activation recomputation~\cite{chen2016trainingdeepnetssublinear, korthikanti2022reducingactivationrecomputationlarge} to further reduce the memory footprint. When scaling out across multiple consumer-grade GPUs connected via PCIe, data parallelism techniques like ZeRO-Infinity~\cite{rajbhandari2021zeroinfinitybreakinggpumemory} distribute model states across devices. However, they require frequent collective communications at each layer to reconstruct parameters. Prior studies~\cite{feng2023mobius} indicate that these communications can consume up to 70\% of the training time.
To mitigate such PCIe bottlenecks, Mobius~\cite{feng2023mobius}, as the state-of-the-art system, integrates pipeline parallelism (PP) with offloading, replacing extensive multi-GPU collectives with more efficient P2P communication. 
Mobius employs a looped-BFS schedule~\cite{lamypoirier2023breadthfirstpipelineparallelism} and supports offloading by loading and evicting model weights at the granularity of individual pipeline stages.

However, the looped-BFS schedule, as well as other pipeline schedules designed for datacenter training, suffers from significant pipeline bubbles.
In these schedules, the weights of a stage (and the corresponding forward and backward computation) are fixed to a specific GPU. We term this the \textit{weight binding} issue. No matter how the pipeline is partitioned (standard looped partitioning in Figure~\ref{fig:rpp-schedule}~(a) or flexible partitioning in Figure~\ref{fig:rpp-schedule}~(b)), GPUs have to wait for the slowest stage (e.g., the LM Head), causing structural bubbles between stages or imbalanced bubbles within stages. In the case of training the Llama-3.1-8B model, pipeline bubbles can reach up to 30\%.

Our core observation is that CPU offloading offers an opportunity to break the \textit{weight binding} issue. In the offload setting, GPUs can be viewed as a stateless execution worker pool. Master weights reside in host DRAM; during the forward pass, the weights of each stage are transferred to a GPU for computation, and the resulting activations are stored back to host DRAM; during the backward pass, the same stage can be dispatched to a different GPU dynamically.

Based on this insight, we propose \rpp{}, whose pipeline schedule is shown in Figure~\ref{fig:rpp-schedule}~(c).
\rpp{} leverages a round-robin task dispatching schedule and assigns stages to different GPUs in a sequential manner.
This allows \rpp{} to freely select an optimal number of pipeline partitions without imbalance bubbles, while existing schedules are limited to multiples of \#GPUs.
To pipeline the forward and backward stages without bubbles, \rpp{} adopts an asymmetric stage splitting strategy to balance execution time across stages.
For example, we combine three layers into a forward stage or one layer into a backward stage.
\footnote{Due to activation recomputation, their execution times are equal.} 
This also removes iteration boundaries, allowing the pipeline to flow from one iteration to the next without structural bubbles when performing gradient accumulations or using asynchronous optimizers.

Fully unleashing the potential of \rpp{} requires tackling several challenges. First, massive amounts of data must continuously move between the host and GPUs. 
To prevent large parameter transfers from blocking critical-path activation transfers and the compute stream, \rpp{} introduces a priority-aware transfer scheduling engine, packing parameter transfers into idle windows between critical-path activation transfers, ensuring seamless overlap with computation.

Second, \rpp{} supports executing optimizer updates asynchronously to hide latency. This introduces race conditions over host-resident data, where the GPU and the CPU-side optimizer simultaneously read and write this data. To guarantee data consistency without reintroducing pipeline-stalling barriers, \rpp{} implements a \lmcp{} that enforces strict execution ordering at the individual layer level. 

Finally, to eliminate manual tuning and maintain near-optimal load balancing across asymmetrically partitioned stages, \rpp{} features an automated stage-splitting algorithm that efficiently (with $O(L^{3})$ complexity) computes a pipeline partition. 

We evaluate \rpp{} across 8$\times$RTX 4090 and 8$\times$A800 SXM servers using 1.7B to 235B models~\cite{qwen3technicalreport,grattafiori2024llama3herdmodels,openai2025gptoss120bgptoss20bmodel}. Compared with state-of-the-art baselines, \rpp{} delivers: (1) on 4090s, up to $2.16\times$ higher throughput and $7.3\times$ longer sequences; (2) on A800s, matching their throughput for small models, with up to $1.47\times$ speedups and $5.6\times$ longer sequences on large ones. Remarkably, \rpp{} is the only system capable of LoRA fine-tuning a 235B MoE model on 24\,GB GPUs. Furthermore, its 4090 throughput reaches at least 76\% of existing A800 solutions across all models, bridging the performance gap between consumer and data-center hardware.

In summary, the main contributions of this paper are as follows:
\begin{compactitem}[\scriptsize{$\bullet$}]
    \item We identify limitations of pipeline parallelism on consumer GPUs and propose \rpp{} as a solution.
    \item We propose a set of system designs, including a multi-stream architecture for data overlap and a \lmcp{}, alongside a stage-splitting algorithm to fully unleash the potential of \rpp{}.
    \item We evaluate \rpp{} comprehensively to demonstrate its effectiveness and training efficiency compared to state-of-the-art solutions.
\end{compactitem}

%% file: sec/background.tex
\section{Background and Motivation}
\label{sec:background}
Consumer-grade GPUs offer high computational power at a fraction of the cost of datacenter accelerators. However, their limited hardware memory capacity and low-bandwidth PCIe interconnects pose bottlenecks for training large models. In this section, we first review the memory pressure and existing mitigation techniques (\S\ref{sec:bg-memory}). We then discuss multi-GPU parallelism under the offloading setting (\S\ref{bg:pp-offload}), highlighting the limitations of current pipeline schedules in resolving structural and imbalance bubbles (\S\ref{sec:bg-pp-schedules}).
\subsection{Memory-Consuming Training and Mitigations}
\label{sec:bg-memory}

On memory-constrained consumer GPUs, one primary challenge in training or fine-tuning large models is memory pressure. This pressure stems from the massive footprint of static \textit{model states} (including parameters, gradients, and optimizer states) and runtime intermediate activations.

First, the continuous scaling of model size to unlock emergent capabilities~\cite{kaplan2020scalinglawsneurallanguage} increases the basic memory required for model states. 
Under standard mixed-precision~\cite{micikevicius2018mixedprecisiontraining} training with the Adam optimizer~\cite{rajbhandari2020zeromemoryoptimizationstraining}, a $\Phi$-parameter model state occupies $16\Phi$ bytes. For a 32B model, the basic memory usage has reached 512GB.
 
Second, the continuous expansion of context windows for complex tasks like reasoning~\cite{Guo_2025}, video generation~\cite{wan2025wanopenadvancedlargescale}, and document-level understanding~\cite{an2023levalinstitutingstandardizedevaluation} drastically inflates the runtime activation that must be stored for backpropagation.
The activation footprint per transformer layer increases linearly with input sequence length.
For example, training a LLaMA-3.1-8B~\cite{grattafiori2024llama3herdmodels} model with a single 16k-token sequence generates $68$ GB of activations from all layers.
Furthermore, the use of micro-batches (which linearly scale up the activation) is common during distributed training.

Several techniques can mitigate the training memory pressure on consumer-grade GPU servers.
We will review representative approaches below and discuss their trade-offs.

\subsubsection{Activation Recomputation}
\label{sec:bg-recompute}

Activation recomputation or gradient checkpointing~\cite{chen2016trainingdeepnetssublinear} avoids storing intermediate activations during the forward pass.
Instead, only the input to each transformer layer is retained. The remaining activations are recomputed on-the-fly by re-executing the forward pass with the stored layer input before the backward pass of every layer.
This reduces the intermediate activations' size to $2sbh$ bytes per layer~\cite{korthikanti2022reducingactivationrecomputationlarge}.

\begin{figure}[t]
\centering
\includegraphics[width=\columnwidth]{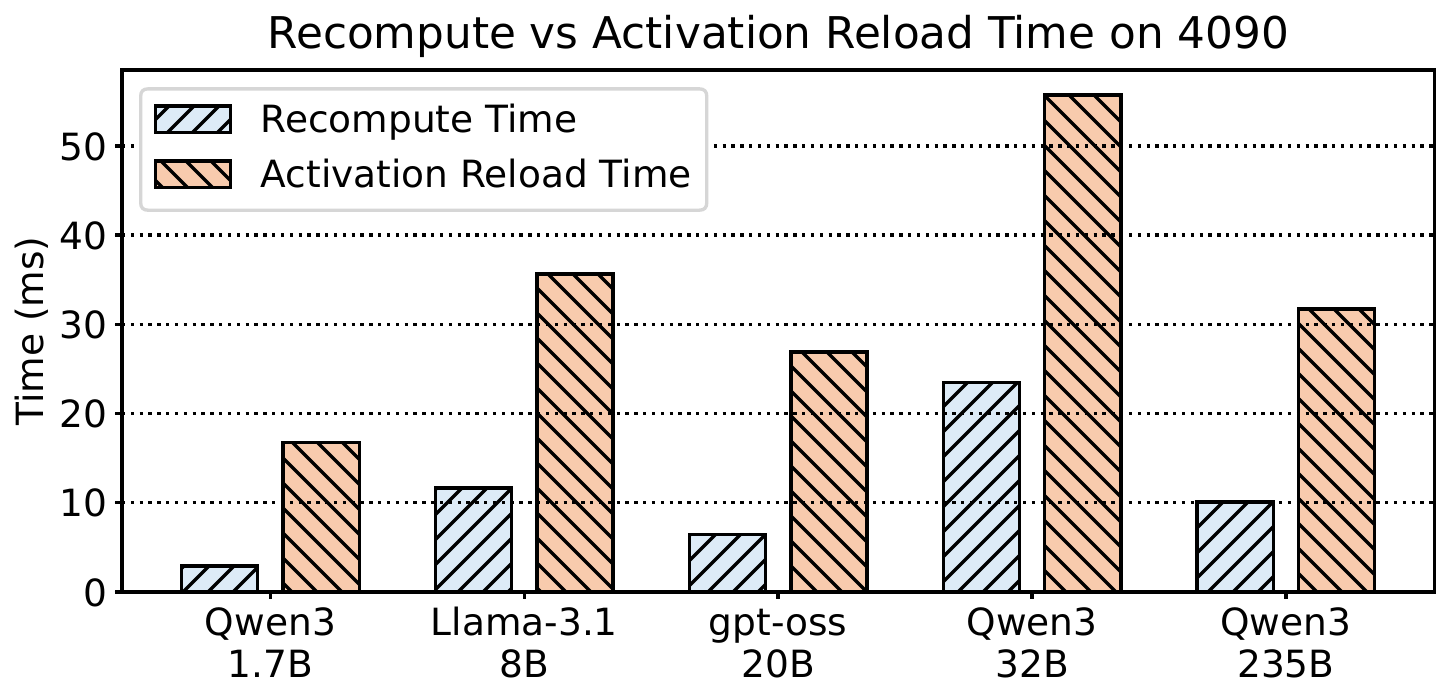}
\vspace{-0.7cm}
\caption{Time of recomputing and reloading activations of a transformer layer.}
\label{fig:act_ckpt}
\end{figure}

While the extra computation may seem expensive, it is often more efficient than swapping activations to host memory over PCIe.
Figure~\ref{fig:act_ckpt} compares the execution time of activation recomputation and reloading activations from host memory on 4090.
Recomputing a transformer layer is $2.37\times \sim 5.75\times$ faster than reloading activations from host memory.
Hence, the reloading overhead cannot even be mitigated by layer-by-layer computation-transmission overlap.
Activation recomputation has therefore become a common practice in large-scale training.
It is favored by mainstream LLM training frameworks~\cite{Zhao_2025}, and is employed even on datacenter clusters with sufficient GPU memory~\cite{jin2025megascalemoelargescalecommunicationefficienttraining,liu2025trainingreporttelechat3moe}.
In this paper, we adopt full activation recomputation as a basic assumption. We also offload the checkpointed activation to host DRAM. 

\subsubsection{Weight and Optimizer Offloading}
\label{sec:bg-optim}

To reduce the static memory footprint of model states, prior work has proposed to offload optimizer states, gradients, and often parameters to host memory, and to execute the optimizer on the CPU~\cite{ren2021zerooffloaddemocratizingbillionscalemodel,rajbhandari2021zeroinfinitybreakinggpumemory}.
Model weights are transmitted to the GPU layer by layer for gradient computation, and the resulting gradients are sent back to the host memory for optimizer updates. A host-side optimizer step is typically slow for large models, 9.6s for a 32B model~\cite{ren2021zerooffloaddemocratizingbillionscalemodel}. Since the execution time of the CPU optimizer is comparable to the scale of GPU computation, synchronous execution of both reduces training efficiency.
Prior work has therefore adopted \emph{staleness-1} asynchronous updates.
Training iteration $T{+}1$ will read the weights produced after iteration $T{-}1$, while the CPU applies iteration-$T$ gradients to model weights in the background~\cite{ren2021zerooffloaddemocratizingbillionscalemodel,lan2025zenflowenablingstallfreeoffloading}.
These works demonstrate that one-step staleness does not harm convergence, making asynchronous optimizer a practical option for memory-constrained training systems.

\subsection{Parallelism with Offloading}
\label{bg:pp-offload}
Given the limited PCIe interconnect of consumer GPUs, another critical factor for training efficiency is how to parallelize across multiple consumer GPUs. 

\paragraph{DP-based parameter offloading}
\label{sec:bg-dp-offload}

A simple approach is to combine data parallelism with host memory offload.
DeepSpeed ZeRO-Offload~\cite{ren2021zerooffloaddemocratizingbillionscalemodel} and ZeRO-Infinity~\cite{rajbhandari2021zeroinfinitybreakinggpumemory} partition model states across data-parallel ranks, and swap them to CPU memory or NVMe storage.

However, as a data-parallel approach, ZeRO requires exchanging the full set of parameters between every GPU for each forward and backward computation. It is usually completed through a high volume of collective communications like \texttt{all-gather}. 
On consumer-grade servers with low PCIe bandwidth, these communications become the major performance bottleneck.
A previous study shows that DeepSpeed spends about 70\% of training time on communication on consumer GPU servers~\cite{feng2023mobius}, limiting the efficiency of multi-GPU scaling on cost-effective hardware.

\paragraph{PP-based parameter offloading}
\label{sec:bg-pp-offload}

To overcome the communication bottleneck inherent in data-parallel offloading, Mobius~\cite{feng2023mobius} proposes a pipeline-parallel approach to offloaded training. It partitions a model's layers into pipeline stages and assigns each stage to a different GPU.
These stages are stored in DRAM, and each stage's weights are loaded to its assigned GPU before executing it. This reduces communication traffic, as only activations and gradients pass between GPUs. However, Mobius inherits the pipeline bubble problem from existing pipeline parallelism.

\subsection{Motivation: Pipeline Schedules and Bubbles}
\label{sec:bg-pp-schedules}

Pipeline parallelism typically suffers from two types of performance-degrading bubbles. As shown in Figure~\ref{fig:rpp-schedule}~(a), 
\emph{Structural bubbles} arise from the data-dependency of forward and backward propagation under balanced stages.
\emph{Imbalance bubbles} arise when some stages run longer than others, forcing dependent stages to stall.
Existing pipeline schedules struggle to mitigate both types of bubbles simultaneously when training large models.

\paragraph{Structural Bubbles.}
Structural bubbles come from data dependencies in pipeline execution.
Non-looped schedules such as GPipe~\cite{huang2019gpipeefficienttraininggiant} and 1F1B~\cite{harlap2018pipedreamfastefficientpipeline} assign one stage to each GPU.
The first stage starts first and finishes last, so some GPUs are idle at the beginning and end of the iteration.Given a model split into $S$ stages and trained on $M$ micro-batches, this results in a pipeline bubble of $\frac{S-1}{M+S-1}$, which can be substantial when $M$ is not much larger than $S$.

Zero-Bubble schedules~\cite{qi2023zerobubblepipelineparallelism} attempt to fill these idle slots by reordering computations. However, they require holding activations for extended periods. This conflicts with activation recomputation and leads to prohibitive memory consumption for large-scale models.
Looped schedules (e.g., Interleaved 1F1B~\cite{narayanan2021efficientlargescalelanguagemodel}, Looped BFS~\cite{lamypoirier2023breadthfirstpipelineparallelism}) offer a more memory-efficient alternative.
They assign an equal number of stages to each GPU, creating a finer-grained pipeline with a total of $S=vN$ stages across all $N$ GPUs.
This reduces the bubble ratio to approximately $\frac{N\cdot (N-1)}{S \cdot M + N\cdot (N-1)}$, which decreases as the number of stages grows.

\paragraph{Imbalance Bubbles.}

\begin{figure}[t]
\centering
\includegraphics[width=\columnwidth]{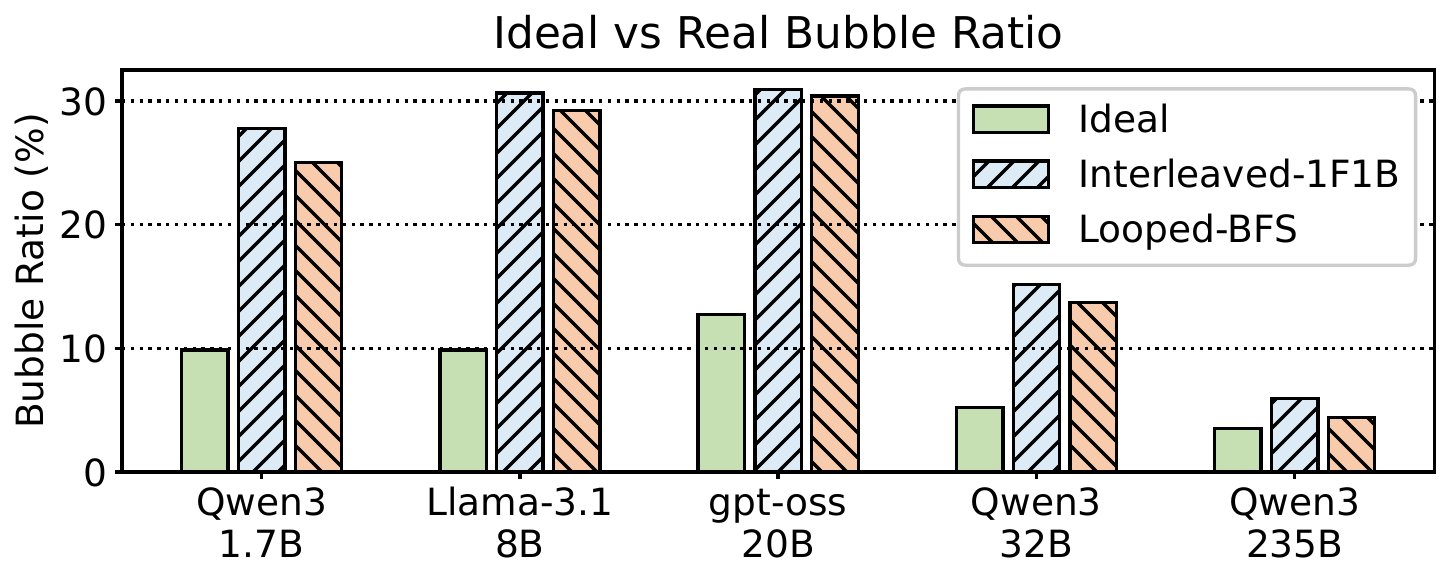}
\caption{Bubble ratio of Looped schedules under ideal balanced partition and real-world imbalanced partition on 8 GPUs. Real-world bubble ratios are collected in \S\ref{sec:eval-ablation-bubble}}
\label{fig:stage-balance}
\end{figure}

Theoretical analyses of pipeline bubbles assume uniform stage execution times, an assumption frequently violated by real-world models.
In practice, stage latencies diverge significantly; the slowest bottleneck stage stalls subsequent operations, inducing imbalance bubbles within the pipeline.
Consequently, this results in a significant disparity between the actual and ideal bubble ratios for the looped schedule, as shown in Figure~\ref{fig:stage-balance}.

\paragraph{Limitations.}
Crucially, existing schedules face a dilemma because they force the total stage count to be an exact integer multiple of the GPU count. Coarse-grained partitioning with fewer stages (e.g., GPipe) incurs high structural bubbles. Conversely, fine-grained partitioning with more stages (e.g., looped schedules) leads to severe load imbalance: as each stage contains fewer layers, compute-heavy components, such as the LM head, introduce significant inter-stage imbalance regardless of where they are assigned. Consequently, as demonstrated in Figure~\ref{fig:stage-balance}, the overall bubble ratio can reach up to 30\% in current pipeline schedules.

%% file: sec/method.tex
\section{Introducing \rpp{}}
\label{sec:method}

In this section, we introduce \rpp{}. By leveraging the opportunities presented by offloading, we decouple pipeline stages from specific physical GPUs and propose a novel \textbf{\cdp{}} (\S\ref{sec:cdp}). Building on this foundation, we employ \textbf{round-robin dispatch} and \textbf{asymmetric splitting} to synthesize a new pipeline schedule supporting an arbitrary number of stages (\S\ref{sec:schedule}). We analyze the bubble ratio of this pipeline and the communication cost of \cdp{} (\S\ref{sec:schedule-analysis}).

\subsection{\cdp{}}
\label{sec:cdp}
\paragraph{Strawman: flexible pipeline partition.}
Recall that existing pipeline schedules require partitioning stages into integer multiples of the number of devices $S=vN$, resulting in imbalance bubbles (as shown in Figure \ref{fig:rpp-schedule}~(a)). An intuitive approach to reduce the imbalance bubble is to apply a flexible stage partitioning (such as 13 stages in Figure~\ref{fig:rpp-schedule}~(b)). This approach can reduce imbalance bubbles by reducing the execution time imbalance between layers.

However, this approach increases structural bubbles. In Figure~\ref{fig:rpp-schedule}~(b), GPU 1 is assigned more stages, while GPUs 2-4 have fewer stages. Unfortunately, the weights (and corresponding computations) of a stage in existing pipeline schedules are fixed to specific GPUs and do not move during training. Therefore, both the forward and backward passes of a stage reside on the same physical device. Since each GPU can only process its pre-assigned and bound stages, GPUs 2-4 are underutilized, creating structural bubbles between their forward and backward passes when waiting for data from the busiest GPU (GPU 1).

\paragraph{Opportunity: computation dispatching.}
\label{sec:cop-shift}

We find that offloading model states to the CPU presents a unique opportunity to increase the pipeline schedule efficiency.
In PP-based offloading, the model states are offloaded to host memory and transferred to the GPU on demand for execution. It is no longer necessary to bind a stage to a fixed GPU.
This allows us to dispatch the stage computation to an underutilized GPU freely.
Intuitively, this incurs no additional communication cost; since weight transfers between the CPU and GPU are already necessary, this method essentially reassigns the target GPU for these weights and their associated computations. We evaluate its cost in \S\ref{sec:schedule-analysis}.

Thus, we propose the \cdp{}.
Model states and activations reside on the host, and computations (with the corresponding model states and activations) are dispatched to GPUs for execution.
In this way, \textbf{any GPU can execute any stage} as soon as the requisite data is ready.
By doing so, we can execute a flexible pipeline with a small structural bubble, achieving a simultaneous reduction in both imbalance and structural bubbles.


\subsection{\rpp{} Schedule}
\label{sec:schedule}

We concretize the idea of computation dispatch with the \rpp{} schedule (Figure~\ref{fig:rpp-schedule}c). \rpp{} incorporates a round-robin stage dispatch pattern and an asymmetric stage splitting strategy.

\paragraph{Round-robin dispatch.} Like looped schedules, \rpp{} distributes stages across GPUs in a round-robin fashion; however, it unifies both forward and backward stages into a single continuous dispatch sequence.
\rpp{} divides the $M$ micro-batches into multiple rounds, processing $M_R$ $(M_R\geq N)$ micro-batches per round on $N$ GPUs.
Within each round, the $S_f$ forward stages are followed by the $S_b$ backward stages, forming a linear sequence of $S = S_f + S_b$ stage slots.
Stage slot $i$ (counting from 0 across the concatenated forward-then-backward sequence) is dispatched to $GPU_{(g_0 + i) \bmod N}$, where $g_0$ is the starting GPU index for the round.
Each GPU executes all $M_R$ micro-batches of the current round for its assigned stage before the next stage slot is dispatched.

Between rounds, the dispatch seamlessly resumes where the previous round left off.
That is, the starting index updates to $g_0 \leftarrow (g_0 + S) \bmod N$, assigning the first stage of the new round to the GPU next in line.
This dispatch logic ensures a continuous, zero-bubble flow of stages across rounds.
Stage dispatch will repeat $R = M / M_R$ rounds until all $M$ micro-batches have completed their forward and backward passes.

\paragraph{Asymmetric stage splitting.}
\label{sec:stage-split}

The forward pass of a transformer layer is faster than its backward pass with recomputation.
The symmetric stage splitting adopted by conventional pipeline parallelism enforces identical layer partitions across both passes, creating idle bubbles at the transition boundary between the faster forward and slower backward stages.
In contrast, \rpp{} maintains separate partitions for forward and backward passes, which enables a balanced partition that equalizes per-stage execution time across the entire forward-to-backward sequence.

To formalize this asymmetric partitioning, \rpp{} defines $(F_1, F_2, \ldots, F_{S_f}, B_1)$ as the forward partition, where $F_i$ is the number of consecutive layers in forward stage $i$; and defines $(B_1, B_2, \ldots, B_{S_b})$ as the backward partition, where $B_j$ is the number of consecutive layers in backward stage $j$.
To reduce redundant computation, \rpp{} fuses the forward and backward passes for the $B_1$ layers.
In this fused stage, these forward computations serve as the recomputation required by the backward pass, saving one forward pass's worth of computation for these layers.
All stages should have approximately equal computation time across both directions and satisfy $\sum_i F_i + B_1 = L, \sum_j B_j = L$, where $L$ is the number of model layers.

By breaking the symmetry between forward and backward partitions, asymmetric splitting bridges the gap between the forward and backward phases, eliminating \rpp{}'s pipeline bubbles at phase boundaries.

\paragraph{Supporting asynchronous optimizer.}
\label{sec:schedule-async}

\begin{figure}[t]
\centering
\includegraphics[width=0.93741\columnwidth]{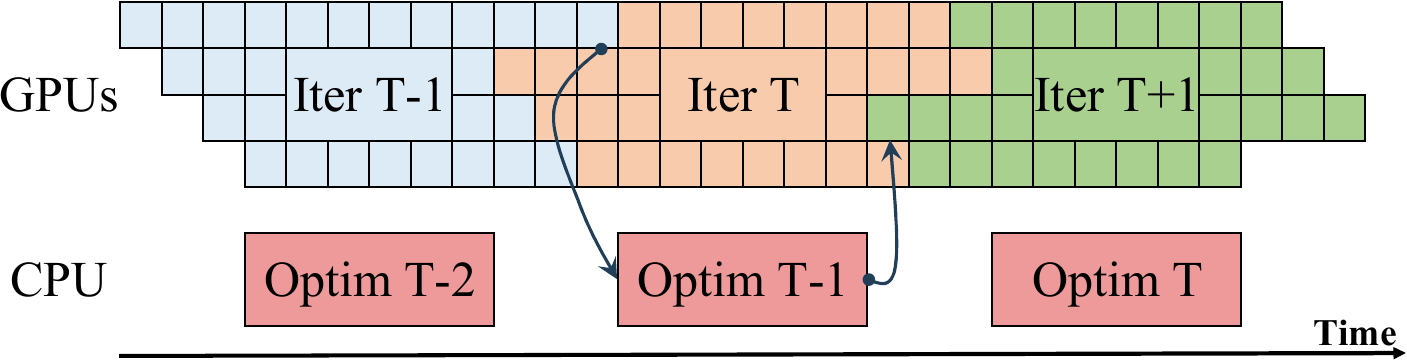}
\caption{Asynchronous optimizer update in \rpp{}.}
\label{fig:rpp-async}
\vspace{-1\baselineskip}
\end{figure}

\rpp{} natively supports staleness-1 asynchronous optimizer update (\S\ref{sec:bg-optim}).
\rpp{}'s round-robin dispatch allows the pipeline to flow from one iteration to the next without structural bubbles.
Figure~\ref{fig:rpp-async} presents how \rpp{} works with an asynchronous optimizer.
The iteration $T$ continues the round-robin assignment from where iteration $T-1$ left off.
With asynchronous updates enabled, the warm-up and cool-down bubbles that appear at iteration boundaries are thereby eliminated.

\subsection{Benefits and Cost Analysis}
\label{sec:schedule-analysis}


\paragraph{Bubble Analysis.}
\label{sec:schedule-bubble}

We now quantify the pipeline bubble ratio of the \rpp{} schedule.
Let $t$ denote the execution time per micro-batch for each stage. With the stage count $S = S_f + S_b$, the workload of all $M$ micro-batches costs $M \cdot S \cdot t$ GPU time.
The bubble of \rpp{} comes from the startup cost of filling the pipeline in the first round and the cool-down cost of draining the pipeline after the last round, which are symmetric and together contribute $N \cdot (N-1) \cdot t$ GPU time.
The total GPU time is therefore $(M \cdot S + N \cdot (N-1)) \cdot t$ and the bubble ratio is $\frac{N \cdot (N-1)}{M \cdot S + N \cdot (N-1)}$.
Although \rpp{} shares the same ratio formula with looped schedules (Interleaved 1F1B, Looped BFS), the number of stages in \rpp{} is the sum of forward and backward stages, which is around $\frac{4}{3}\times$ larger than looped schedules.
\footnote{
To balance forward and backward stage runtime, \rpp{} typically assigns $3$ layers to $1$ forward stage.
Looped schedules have at most \#layers ($L$) stages, while \rpp{} has $\frac{L}{3}$ forward stages and $L$ backward stages ($\frac{4}{3}L$ in total).}
Thus, the bubble ratio of \rpp{} is smaller than that of looped schedules. In addition, \rpp{}'s flexible stage partitioning allows it to achieve better time balance and therefore smaller imbalance-induced bubbles than looped schedules.

\paragraph{Communication Cost.}
\label{sec:cop-roofline}

\begin{figure}[t]
\centering
\includegraphics[width=0.87508\columnwidth]{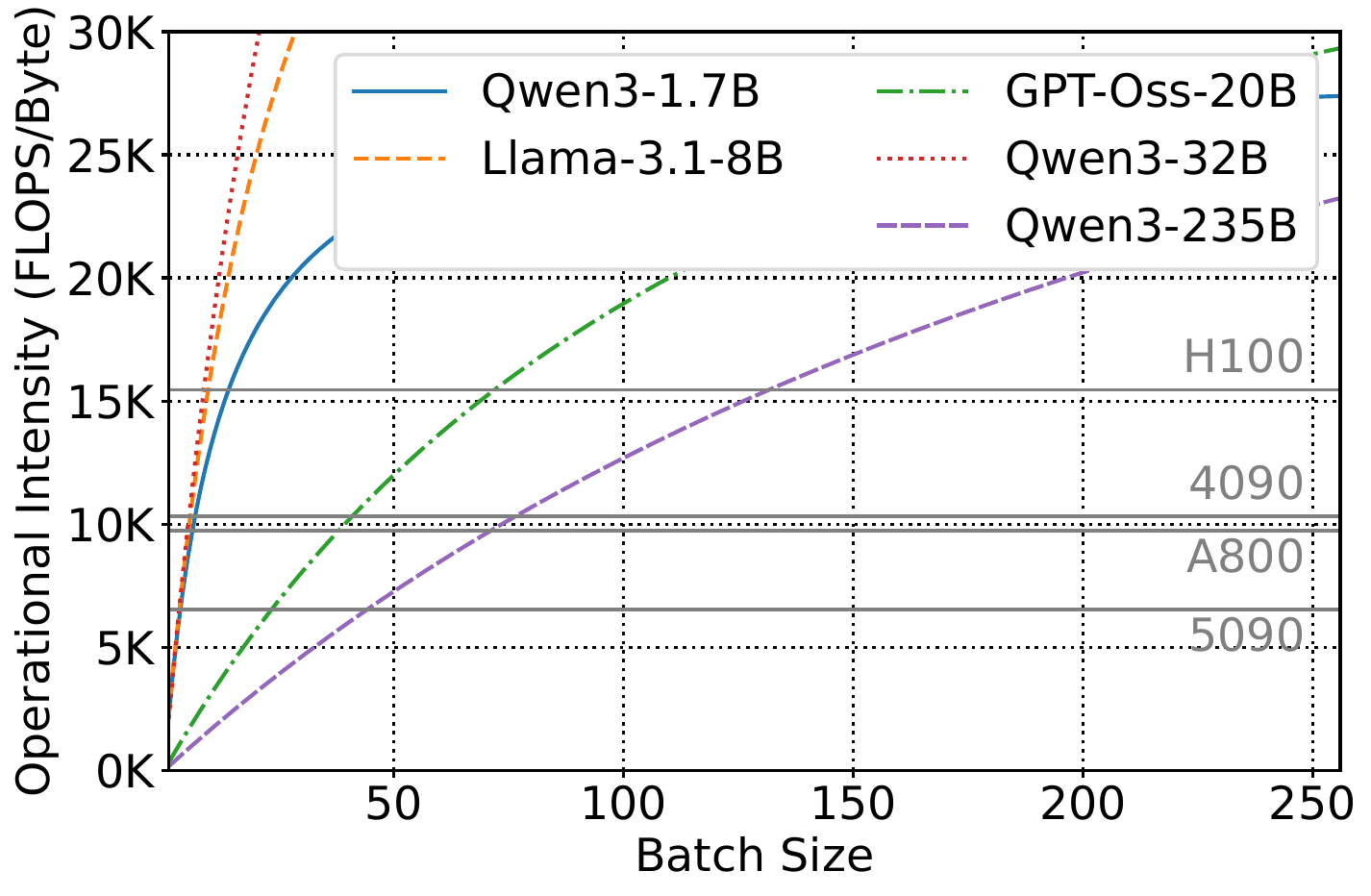}
\caption{Operational intensity vs.\ batch size for representative dense and MoE models at $2048$ sequence length. Horizontal lines mark the ridge-point OI for each GPU.}
\label{fig:oi-vs-batch}
\end{figure}

\rpp{} employs \cdp{} to dispatch stage computations to GPUs, thereby requiring transferring weights and activations between the GPU and the CPU memory.
We conduct a roofline analysis~\cite{williams2009roofline} to evaluate the PCIe communication cost during this dispatch process.

Figure~\ref{fig:oi-vs-batch} presents the operational intensity (OI) of the forward pass within a single transformer layer for multiple models.
When the OI of a workload exceeds the GPU's ridge point, it indicates that the workload is compute-bound, meaning that data transfer latency can be fully hidden by computation.
For dense models, OI exceeds the ridge point at batch size $B=8$, the smallest batch size for 8-GPU pipeline parallel training.
For MoE models, the ridge point is crossed when $B < 100$, well within the typical training regime.
Furthermore, the backward pass (with activation recomputation) has ${\sim}3\times$ the forward FLOPS but ${\sim}2\times$ the data movement, so its OI is higher than the forward pass's.

This proves that model training under \cdp{} is compute-bound.
Therefore, \cdp{} is a throughput-preserving method that unlocks a larger scheduling search space and scales up trainable model parameters and sequence lengths.
Based on this, \rpp{} implements an efficient communication-computation overlap (\S\ref{sec:data-overlap}) and verifies its minimal overhead nature through our evaluation in \S\ref{sec:eval-a800}.

%% file: sec/design.tex
\section{Design and Implementation}
\label{sec:design}

The previous section presented \rpp{}'s scheduling algorithm in the abstract.
Realizing this abstraction as a practical training framework involves mapping the logical pipeline to physical hardware efficiently.
\S\ref{sec:overview} introduces the overall system architecture and workflow, followed by the core technical challenges.
Subsequent subsections detail how \rpp{} addresses data transfer overlap (\S\ref{sec:data-overlap}), parameter consistency (\S\ref{sec:consistency}), and stage partitioning (\S\ref{sec:auto-split}).

\subsection{\rpp{} Overview}
\label{sec:overview}

\begin{figure}[t]
\centering
\includegraphics[height=120.94bp]{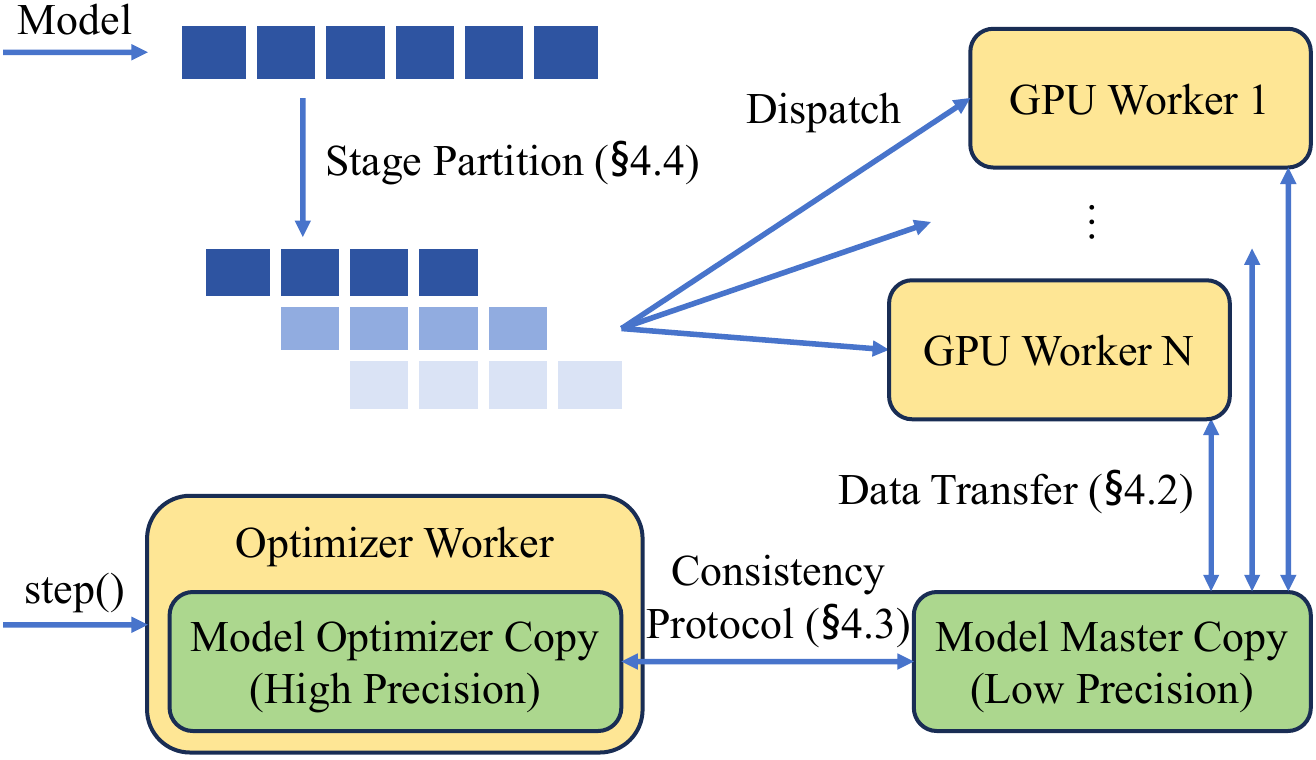}
\caption{\rpp{} system overview.}
\label{fig:overview}
\end{figure}

\rpp{} adopts a single-controller architecture inspired by Ray~\cite{moritz2018raydistributedframeworkemerging} and veRL/HybridFlow~\cite{Sheng_2025}, separating the control plane, which manages task scheduling and ordering, from the data plane, which handles hardware-level execution and device data transfers.
As shown in Figure~\ref{fig:overview}, \rpp{} consists of a controller, an optimizer worker, and one GPU worker for each GPU.
The \rpp{} controller transparently maps the computation onto the available hardware via GPU workers, and handles asynchronous optimizer updates on the optimizer worker.

\subsubsection{\rpp{} Workflow}
\label{sec:workflow}

Users write a sequential training script using two APIs.
During each \rpp{} invocation, the user's thread acts as the controller to orchestrate all activity, directing per-device GPU workers and the optimizer worker that execute concurrently.

\paragraph{forward\_backward()}
From the caller's perspective, this is identical to a standard PyTorch forward and backward pass.
It accepts input tensors, returns a scalar loss, and accumulates gradients in place.
Internally, \rpp{} runs a complete pipelined execution across all available GPUs.
The controller constructs micro-batches from the inputs and dispatches the stage to a GPU worker in round-robin order.
The GPU workers execute in parallel with the controller and with one another, handling the data movement and computation asynchronously.

\paragraph{step()}
This API dispatches gradient post-processing and weight updates (e.g., gradient scaling, clipping, and the optimizer step) to the optimizer worker. 
The optimizer worker applies these updates to the full-precision \textit{optimizer copy} of the model, while the GPU workers concurrently compute the next iteration using the \textit{master copy} without interference.

\subsubsection{Challenges}
\label{sec:challenges}

\begin{figure}[t]
\centering
\includegraphics[width=\columnwidth]{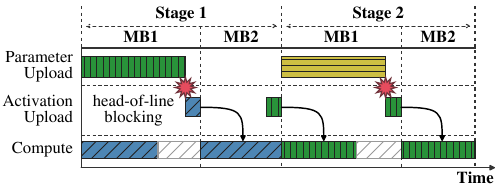}
\caption{Illustrated example of simple compute-transfer overlap on two consecutive stages with two microbatches (MB) each. Each color represents one stage.}
\label{fig:naive-overlap}
\end{figure}

There are several software-design challenges that, if handled carelessly, would introduce subtle performance and correctness bugs:

\begin{compactitem}[\scriptsize{$\bullet$}]
    \item \textbf{Data Transfer Overlap:} Under the \cdp{}, multiple kinds of data move between the host and GPUs continuously.
    As shown in Figure~\ref{fig:naive-overlap}, a simple overlap strategy suffers from head-of-line blocking, where large parameter/gradient transfers delay the critical-path activation transfers for the current stage.
    \S\ref{sec:data-overlap} addresses this via a priority-aware transfer scheduling engine that packs non-critical data into the idle windows of critical-path communications.
    \item \textbf{Asynchronous Consistency:} Operating concurrently on the \textit{optimizer copy} and the \textit{master copy} of the model requires bridging the gap between them.
    The system must propagate updates between these two copies to preserve staleness-1 asynchronous optimizer update semantics without stalling pipeline execution.
    \S\ref{sec:consistency} addresses this via a \lmcp{}.
    \item \textbf{Pipeline Stage Partition:} To balance the load across the pipeline without manual tuning, the system requires an efficient mechanism for partitioning layers automatically.
    However, a naive search for the optimal partition incurs an unacceptable exponential time complexity.
    \S\ref{sec:auto-split} addresses this via an $O(L^3)$ two-stage partitioning algorithm.
\end{compactitem}

\subsection{Data Transfer Overlap}
\label{sec:data-overlap}

\begin{figure}[t]
\centering
\includegraphics[width=\columnwidth]{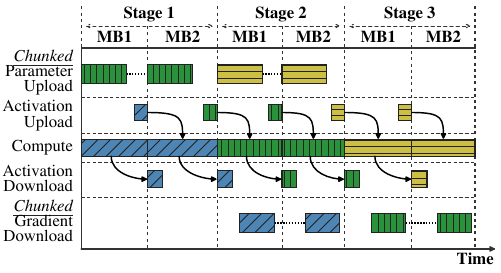}
\caption{\rpp{} multi-stream workflow over three consecutive stages with two microbatches (MB) each. Each color represents one stage.}
\label{fig:streams}
\end{figure}

Under the \cdp{}, every pipeline stage requires two categories of data movement: (1) per-micro-batch activation uploads and downloads between host and GPU, and (2) per-stage parameter uploads and gradient downloads between host and GPU. 

Category (1) lies on the critical path.
The next stage cannot begin until the current stage's output activation has arrived in host memory and been re-uploaded to the target GPU.
Category (2), in contrast, has lower priority and can be scheduled into any available execution window during the preceding or subsequent stages.

\subsubsection{The Multi-Stream Architecture.}
To exploit this overlap and fully utilize the bidirectional PCIe link, \rpp{} maintains four dedicated communication streams per device alongside the default compute stream.
These four streams independently handle the two transfer categories in both directions (upload and download).
Figure~\ref{fig:streams} illustrates this overlap.
Activation uploads and downloads are performed one micro-batch early/delayed to avoid blocking compute, while parameter uploads and gradient downloads are interleaved into idle intervals between activation transfers.
The compute stream synchronizes with the activation transfer streams at micro-batch granularity using CUDA events.
This divides the timeline into bounded data-transfer windows, ensuring dependencies are met while capping in-flight data to avoid out-of-memory errors.

\begin{figure*}[t]
\centering
\includegraphics[width=475.783bp]{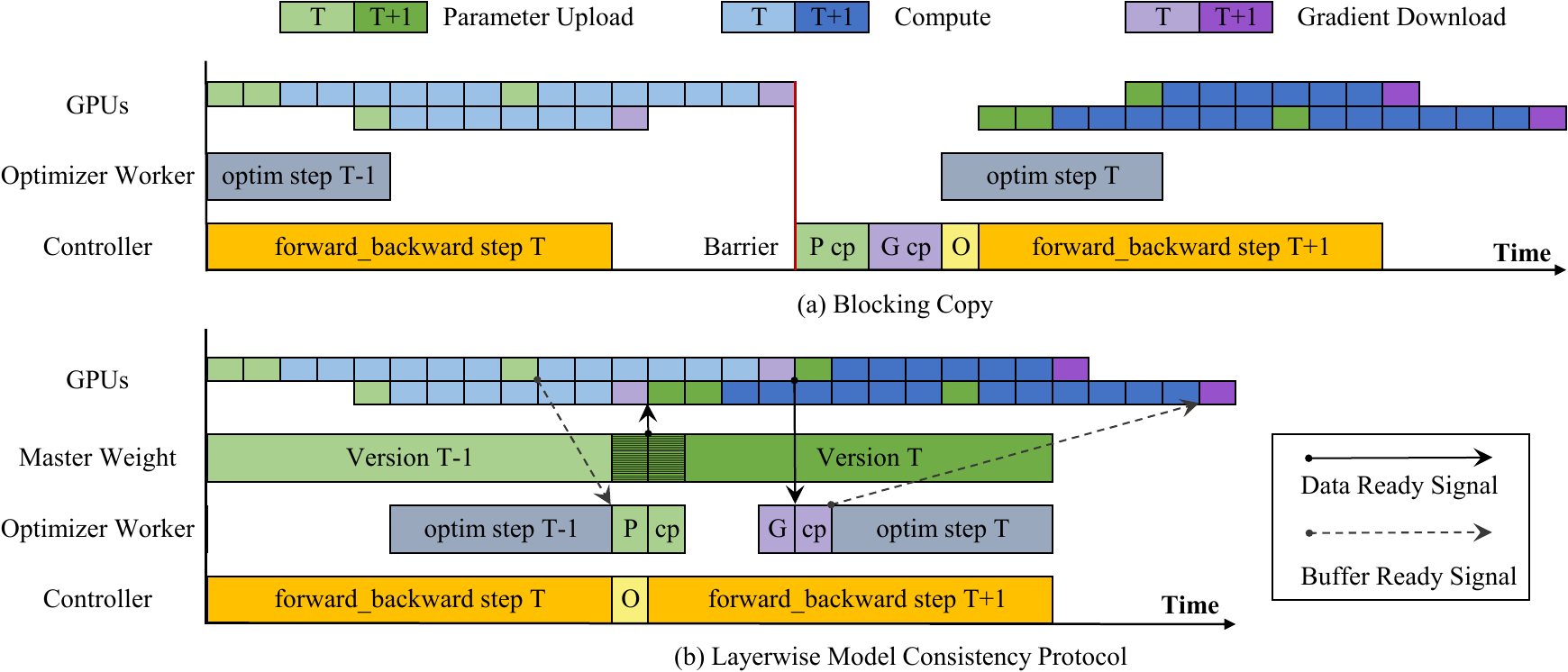}
\caption{
(a) The blocking approach copies weights (\textit{P~cp}) and gradients
(\textit{G~cp}) on the main thread.
(b) \rpp{} offloads copies to the optimizer worker (O) and applies \lmcp{} to ensure correctness.
The four arrows from left to right correspond to ordering constraints (1)-(4), respectively.}
\label{fig:consistency}
\vspace{-0.5\baselineskip}
\end{figure*}

\subsubsection{Scheduling Parameter Transfers.}
To avoid bandwidth contention on the shared PCIe link, \rpp{} confines parameter and gradient transfers to the idle intervals between activation transfers.
A stage processing $M$ micro-batches creates $M$ data-transfer windows, each processing one micro-batch activation and one chunk of parameter/gradient transfers.
\rpp{} partitions parameter/gradient transfers into $M$ chunks by applying longest-processing-time-first scheduling~\cite{Graham1969Bounds}.
Specifically, we sort the parameter/gradient tensors by size in descending order and assign each tensor to the interval with the smallest current total assigned transfer size.
In case of very large tensors (e.g., language model head), we split them into smaller chunks before scheduling to ensure they can fit within the available data-transfer windows without causing contention.


\subsection{Fine-Grained Model Consistency}
\label{sec:consistency}

\subsubsection{The Consistency Problem}
Mixed-precision training with CPU offloading maintains three representations of the parameters and gradients~\cite{ren2021zerooffloaddemocratizingbillionscalemodel}: (1) a short-lived GPU transient copy for active computation, (2) a low-precision CPU master copy in host memory, and (3) a full-precision optimizer copy synchronized with the master.

To coordinate iterations, \rpp{} utilizes the \texttt{step()} call as a logical barrier during asynchronous optimizer updates. 
\rpp{} must precisely control the synchronization of updated weights into the master copy (\textit{P~cp}) and the extraction of gradients from the master copy for the optimizer (\textit{G~cp}).
To guarantee staleness-1 training semantics, correctness reduces to enforcing five ordering constraints among these synchronization operations, GPU data transfers, and optimizer steps:
\begin{compactenum}[(1)]
\item \textbf{Protects against early overwriting of weights.}
    \textit{P~cp} must wait until the GPU has finished uploading parameters for the previous iteration.
\item \textbf{Ensures complete weight updates.}
    The GPU must wait until \textit{P~cp} completes before uploading parameters for the next iteration.
\item \textbf{Protects against reading incomplete gradients.}
    \textit{G~cp} must wait until the GPU has finished downloading the previous iteration's gradients into the master buffer.
\item \textbf{Protects previous gradients from being overwritten.}
    The GPU must wait until \textit{G~cp} completes before writing the next iteration's gradients into the master buffer.
\item \textbf{Preserves optimizer step semantics.}
    Data copies must be scheduled between the just-completed optimizer step and the next incoming step.
\end{compactenum}

\subsubsection{Blocking vs. \lmcp{}}
\label{sec:consistency-protocol}

A straightforward blocking copy (Figure~\ref{fig:consistency}(a)) treats the \texttt{step()} call as a concrete barrier, blocking until all workers have finished their tasks before performing data synchronizations (\textit{P~cp} and \textit{G~cp}).
It is trivially correct, but it reintroduces exactly the pipeline bubble from pipeline flush that asynchronous execution was designed to eliminate.
The data synchronization itself is also time-consuming as it requires performing floating-point casting on the CPU.

To eliminate this, \rpp{} uses a \textbf{\lmcp{}} (Figure~\ref{fig:consistency}(b)).
\rpp{} offloads \textit{P~cp} and \textit{G~cp} to the optimizer worker and coordinates access through a set of threading events.
The ordering constraints (1)-(4) should be preserved by four dependency edges, as shown in Figure~\ref{fig:consistency}(b), while (5) is naturally enforced by the optimizer worker's internal sequential execution.
\rpp{} utilizes point-to-point event signaling to enforce these edges.
The controller dynamically creates and dispatches events for workers to wait for or to set, allowing them to resolve ordering constraints directly in a peer-to-peer manner.

Crucially, synchronization applied to the whole model would still stall the shallowest layer (Layer 1), which sits at the tight turnaround between the end of the backward pass and the beginning of the next forward pass.
Therefore, \rpp{} binds these synchronization events to individual layers.
This fine-grained protocol allows the optimizer worker to release events for early layers immediately after processing them, empowering GPU workers to begin the next iteration's forward pass on Layer 1 while deeper layers are still synchronizing.


\subsection{Automatic Stage Partitioning}
\label{sec:auto-split}

Optimal stage partitioning is critical to maximizing the execution performance of \rpp{}.
All stages should have similar execution times to ensure smooth and bubble-free round-robin stage dispatching.
For the first few iterations, \rpp{} runs a one-layer-per-stage partition to collect runtime statistics.
\rpp{} uses lightweight CUDA events to collect per-layer execution times and tracks memory consumption through the PyTorch allocator.
Based on these statistics, \rpp{} computes a well-balanced stage partition and applies it to subsequent iterations.

\subsubsection{Problem Formulation}

The objective of the partitioning algorithm is to minimize pipeline execution time while ensuring that the partition does not exceed memory capacity.
We model execution time using per-layer computation latency, given that \rpp{} effectively overlaps most communication with computation.
As discussed in \S\ref{sec:schedule-bubble}, total GPU time (including startup and cool-down cost) for \rpp{} to process $M$ micro-batches on $N$ GPUs with $S$ stages is $M\cdot S\cdot t + N \cdot (N-1) \cdot t$ under a balanced partition.
In practice, however, the execution time varies across stages.
To simplify, we model the execution time of all stages as the maximum stage runtime (i.e., the pipeline bottleneck).
Our goal is to minimize $(M\cdot S + N \cdot (N-1)) \cdot t_{\max}$ (\S\ref{sec:schedule-bubble}) while preserving GPU memory constraints on every stage.

\subsubsection{Algorithm}

\rpp{} leverages the observation that, since each stage must consist of contiguous layers, the set of possible values for $t_{\max}$ is bounded to all contiguous sub-sequence sums of forward and backward times ($O(L^2)$ possibilities).
For each candidate $t_{\max}$ value, the problem reduces to: find the minimum number of contiguous partitions such that no partition's total time exceeds $t_{\max}$ and no partition's memory consumption exceeds the GPU's memory limit.
This is a classic greedy problem solvable in $O(L)$ time by scanning layers and packing them into stages until a constraint is violated, resulting in an overall $O(L^3)$ complexity for the partitioning algorithm.
During the greedy process, \rpp{} prioritizes filling the first backward stage to its maximum capacity.
This greedy choice maximizes the size of the first backward stage, which is beneficial because this stage skips activation recomputation (\S\ref{sec:stage-split}), leading to greater reductions in total compute cost.

%% file: sec/eval.tex
\section{Evaluation}
\label{sec:eval}

In this section, we first evaluate \rpp{} on both consumer-grade and datacenter-grade GPU servers, then quantify its host memory consumption and training cost, and study its scaling with GPU count and sequence length.
We finally break down the pipeline bubbles of \rpp{} and evaluate the effectiveness of \rpp{}'s design.

\subsection{Experimental Setup}
\label{sec:eval-setup}

\paragraph{Hardware}
We use two server configurations. Unless otherwise stated, all experiments use 8 GPUs on a single server.
\begin{compactitem}[\scriptsize{$\bullet$}]
\item \textbf{4090 server:} 8$\times$ NVIDIA RTX~4090 GPUs (24\,GB VRAM each), Intel Xeon Gold 6330 CPU, 800\,GB available DDR4 host memory, PCIe~4.0 (32\,GB/s) interconnect.
\item \textbf{A800 server:} 8$\times$ NVIDIA A800 SXM GPUs (80\,GB HBM2e each), Intel Xeon Platinum 8352Y CPU, 800\,GB available DDR4 host memory, NVLink~3.0 (200\,GB/s) interconnect.
\end{compactitem}

\paragraph{Baselines}
We compare against six representative training frameworks:
\begin{inparaenum}[(1)]
\item DeepSpeed ZeRO-2~\cite{rajbhandari2020zeromemoryoptimizationstraining},
\item PyTorch FSDP~\cite{paszke2019pytorchimperativestylehighperformance} (ZeRO-3 data parallelism),
\item DeepSpeed ZeRO-Infinity~\cite{rajbhandari2021zeroinfinitybreakinggpumemory} (CPU memory offloading),
\item Megatron-LM pipeline parallelism (Megatron-PP)~\cite{narayanan2021efficientlargescalelanguagemodel},
\item Megatron-LM tensor parallelism (Megatron-TP)~\cite{shoeybi2020megatronlmtrainingmultibillionparameter}, and
\item Mobius~\cite{feng2023mobius}.
\end{inparaenum}
Megatron-PP implements the interleaved-1F1B pipeline schedule, and Mobius implements the looped-BFS pipeline schedule with CPU memory offloading.
We also report \rpp{}-sync, a variant of \rpp{} that disables asynchronous optimizer updates (\S\ref{sec:schedule-async}) and performs a synchronous CPU optimizer step.

\paragraph{Workloads}
We evaluate five open-source models: Qwen3-1.7B~\cite{qwen3technicalreport}, LLaMA-3.1-8B~\cite{grattafiori2024llama3herdmodels}, GPT-OSS-20B~\cite{openai2025gptoss120bgptoss20bmodel}, Qwen3-32B, and Qwen3-235B-A22B.
GPT-OSS-20B and Qwen3-235B-A22B are Mixture-of-Experts (MoE) models, while the other three are dense Transformers.
We perform full-parameter training on the first four models and apply LoRA fine-tuning ($r=32$) exclusively to the massive Qwen3-235B.
The sequence length is fixed at 2048, with global batch sizes of 512, 256, 128, 128, and 64 for each model, respectively.
All frameworks employ mixed-precision (FP16) training, full activation recomputation, and identical global batch sizes.
We maximize the micro-batch size for each framework and model individually to ensure full GPU utilization, subject to two constraints: avoiding device out-of-memory (OOM) and maintaining a sufficient number of micro-batches (e.g., $\geq 8$) to sustain high pipeline or data parallel throughput.

\paragraph{Metrics}
We report two metrics:
(1) Training throughput (tokens/s): the number of tokens processed per second during steady-state training, averaged over 10 iterations after warm-up.
(2) Maximum sequence length: the longest sequence length at which training completes without out-of-memory errors, searched at micro-batch size=1.

\subsection{End-to-End Performance on 4090}
\label{sec:eval-4090}

\begin{figure}[t]
\centering
\includegraphics[width=\columnwidth]{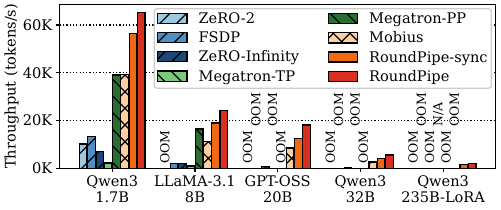}
\caption{Training Throughput on 8$\times$RTX 4090.}
\label{fig:4090-throughput}
\end{figure}

\begin{figure}[t]
\centering
\includegraphics[width=\columnwidth]{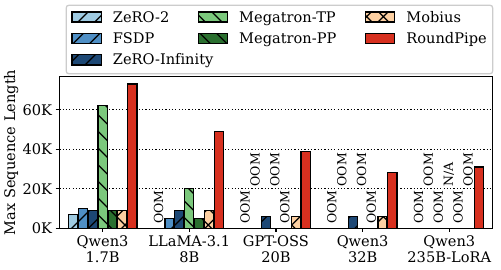}
\caption{Maximum trainable sequence length on 8$\times$RTX 4090.}
\label{fig:4090-seqlen}
\end{figure}

Figures~\ref{fig:4090-throughput} and~\ref{fig:4090-seqlen} present end-to-end throughput and maximum sequence length results on the 4090 server.
\footnote{N/A on Megatron-TP when LoRA finetuning Qwen3-235B because the model has four key value heads, which does not support TP=8 on Megatron.}

\paragraph{Throughput}
Both \rpp{} and its synchronous variant achieve the highest throughput across all five models.
\rpp{} outperforms the fastest existing systems across 1.7-32B models by $1.48\sim 2.16\times$ on training throughput, and \rpp{}-sync outperforms by $1.15\sim 1.63\times$.
This performance advantage is driven by our novel pipeline design, which substantially mitigates pipeline bubbles and yields improved communication-computation overlap compared with existing frameworks.
Furthermore, by strictly bounding the number of layers resident on the GPU, \rpp{} is the only system capable of successfully LoRA-finetuning the massive 235B model on 24GB GPUs.

\paragraph{Maximum Sequence Length}
\rpp{} consistently supports the longest sequences.
Excluding Megatron-TP, whose throughput becomes impractical under PCIe, \rpp{} extends the maximum sequence length by $4.7\sim 7.3\times$ over the next-best baseline.
Optimizer asynchrony does not affect this metric because it does not change GPU memory usage.
On Qwen3-1.7B, \rpp{} and Megatron-TP both handle substantially longer sequences than other baselines.
TP achieves this by sharding activations across GPUs, while \rpp{} achieves it by storing stage-boundary activations in host memory and recomputing layer-internal activations on demand.
\rpp{} avoids TP's heavy communication overhead, thereby achieving higher training throughput than TP.
As model size scales up, non-offloading systems are bottlenecked by model footprint and fail to scale sequence lengths, whereas existing offloading systems (e.g., ZeRO-Infinity) still OOM early because they do not offload activations.
\rpp{} even supports a slightly longer sequence on Qwen3-235B than on Qwen3-32B because LoRA fine-tuning produces fewer intermediate activations and gradients than full-parameter training during recompute and backward pass.

\subsection{End-to-End Performance on A800}
\label{sec:eval-a800}

\begin{figure}[t]
\centering
\includegraphics[width=\columnwidth]{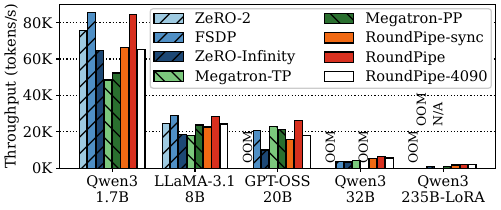}
\caption{Training Throughput on 8$\times$A800.}
\label{fig:a800-throughput}
\end{figure}
\begin{figure}[t]
\centering
\includegraphics[width=\columnwidth]{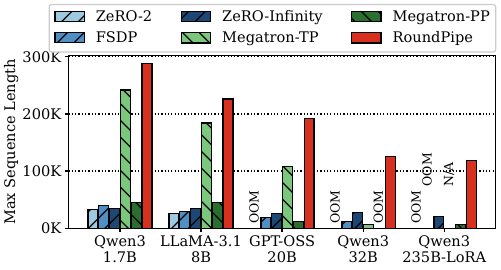}
\caption{Maximum trainable sequence length on 8$\times$A800.}
\label{fig:a800-seqlen}
\end{figure}

To evaluate the overhead of \rpp{} and the \cdp{}, we compare \rpp{} with SOTA training systems on datacenter GPUs with ample GPU memory and interconnect.
Figures~\ref{fig:a800-throughput} and~\ref{fig:a800-seqlen} present the results.
We omit Mobius because it reports lower performance on datacenter GPUs compared to DeepSpeed ZeRO-Infinity~\cite{feng2023mobius}.

\paragraph{Throughput}
Even on hardware-rich servers, \rpp{} delivers $0.98\sim 1.47\times$ the throughput of SOTA systems.
For smaller models (1.7B and 8B), Data Parallelism (DP) performs best by fully exploiting the high-bandwidth NVLink.
Remarkably, \rpp{} achieves highly competitive performance without utilizing any GPU peer-to-peer communication (NVLink), relying entirely on PCIe host-to-device transfers under the \cdp{}.
Meanwhile, \rpp{}-sync matches the speed of DeepSpeed ZeRO-Infinity, with their slowdown attributed to CPU optimizer step overhead.
For larger models (20B+), TP and PP overtake DP by avoiding full-parameter synchronization, but \rpp{} still leads by generating less communication than DP/TP and fewer bubbles than existing PP systems.
On Qwen3-32B, Megatron-PP encounters an OOM error because the 64-layer model must be partitioned across 8 GPUs.
The final rank is burdened with 8 layers plus the LM head, encountering OOM during backward.
Furthermore, Figure~\ref{fig:a800-throughput} plots \rpp{}'s performance on the 4090 server.
Although the 4090's lower VRAM bandwidth inherently caps its peak compute utilization compared to the A800, \rpp{} still consistently achieves over 76\% of the throughput of existing SOTA frameworks running on the A800 across all models.
This proves that \rpp{} elevates consumer GPUs beyond cost-effective alternatives; they now offer absolute training times comparable to the expensive datacenter hardware.

\paragraph{Maximum Sequence Length}
On the A800 server, \rpp{} increases the maximum sequence length by $1.19\sim 5.62\times$ over the baselines, following the same trend as on the 4090.
For smaller models ($\leq 20$B), Megatron-TP uses NVLink to support 108K$\sim$242K tokens, but \rpp{} pushes this further to 192K$\sim$288K by storing model states and stage-boundary activations in CPU memory.
For larger models, TP and the other baselines again become limited by GPU memory, whereas \rpp{} maintains its lead.

\begin{revblock}
\subsection{Host Memory Consumption}
\label{sec:eval-hostmem}

\begin{table}[t]
\centering
\revcolor
\caption{\rev{Host memory consumption (GB) of \rpp{}, \textit{Min}imum required and \textit{Peak} during the benchmark of \S\ref{sec:eval-4090}.}}
\label{tab:hostmem}
\small
\setlength{\tabcolsep}{4pt}
\begin{tabular}{lrrrrr}
\toprule
 & Qwen3 & LLaMA & GPT-OSS & Qwen3 & Qwen3 \\
 & 1.7B & 3.1-8B & 20B & 32B & 235B-LoRA \\
\midrule
Min. & 45.6 & 160.8 & 401.6 & 652.5 & 502.4 \\
Peak & 74.3 & 204.4 & 446.1 & 752.3 & 750.6 \\
\bottomrule
\vspace{-0.1cm}
\end{tabular}
\end{table}

\rpp{} consumes a fixed amount of host memory, approximately $20\times$ the model parameter size.
This footprint includes the FP32 model weights, gradients, and optimizer states, the low-precision weights and gradients, and a small allocation for the intermediate activation tensors used during recomputation.
We monitored host memory utilization during the evaluation of \S\ref{sec:eval-4090}.
As shown in Table~\ref{tab:hostmem}, \rpp{} achieves its peak training throughput and supports its maximum sequence length as long as the server is equipped with host memory matching the minimum requirement.
\end{revblock}

\subsection{Cost Efficiency}
\label{sec:eval-cost}

\begin{figure}[t]
\centering
\includegraphics[width=\columnwidth]{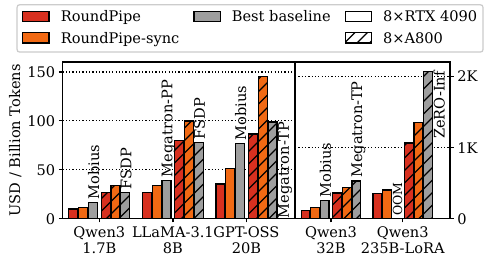}
\par\vspace{-0.1cm}
\caption{\rev{Training cost in USD per billion tokens.}}
\vspace{-0.1cm}
\label{fig:cost-efficiency}
\end{figure}

\begin{revblock}
The 8$\times$RTX~4090 server used in our experiments is priced at \$2.32 per hour, whereas the 8$\times$A800 server costs \$8.16 per hour.
Figure~\ref{fig:cost-efficiency} converts the throughput of \S\ref{sec:eval-4090} and \S\ref{sec:eval-a800} into the training cost per billion tokens.
On the 4090 server, \rpp{} reduces training costs by 30\%--53\% compared to the best existing baselines, and it is cheaper than all systems on the A800 server for every model.
\end{revblock}

\subsection{Scalability}
\label{sec:eval-scale}

\begin{figure}[t]
\centering
\includegraphics[width=190bp]{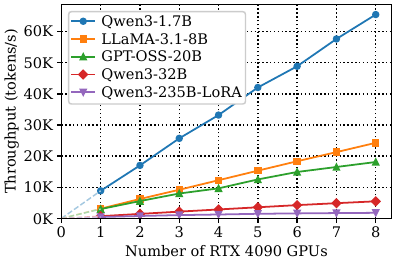}
\par\vspace{-0.1cm}
\caption{\rpp{} throughput on 1$\sim$8 RTX~4090 GPUs.}
\vspace{-0.1cm}
\label{fig:scalability}
\end{figure}

\begin{figure*}[t]
\centering
\begin{minipage}[t]{0.32\textwidth}
\centering
\includegraphics[height=3.2cm]{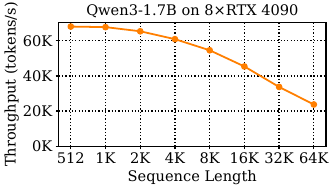}
\caption{Throughput vs. sequence length for Qwen3-1.7B. }
\label{fig:seqlen}
\end{minipage}\hfill
\begin{minipage}[t]{0.32\textwidth}
\centering
\includegraphics[height=3.2cm]{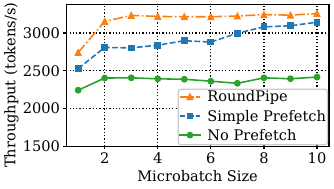}
\caption{Throughput vs. microbatch sizes for Llama3.1-8B on RTX~4090.}
\label{fig:abl-transfer}
\end{minipage}\hfill
\begin{minipage}[t]{0.32\textwidth}
\centering
\includegraphics[height=3.2cm]{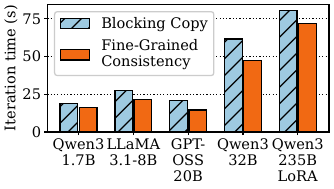}
\caption{Blocking copy vs. \lmcp{}.}
\label{fig:copy-overhead}
\end{minipage}
\end{figure*}

We conducted strong-scaling experiments with fixed workloads to evaluate \rpp{}'s multi-GPU efficiency.
Figure~\ref{fig:scalability} illustrates that \rpp{} achieves near-linear throughput scaling from 1 to 8 GPUs across all model sizes.
The stability of this scaling curve underscores the robustness of \rpp{}, showing that its partitioning and overlap mechanisms remain effective across GPU counts.

Crucially, \rpp{} exhibits a unique architectural advantage: the \textbf{maximum sequence length is independent of the GPU count}.
From 1 to 8 GPUs, the maximum sequence length remains 73K, 49K, 39K, 28K, and 31K for the five models.
This invariant is a direct product of our \cdp{}.
A GPU holds data required for its currently assigned stage only, while the remaining model state and stage-boundary tensors for recompute are stored in host memory.

\subsection{Sequence Length Sensitivity}
\label{sec:eval-seqlen}

Figure~\ref{fig:seqlen} shows how \rpp{}'s throughput changes as the sequence length increases over two orders of magnitude on 8$\times$RTX 4090 with Qwen3-1.7B.
\rpp{} maintains robust performance across the entire range, with throughput decreasing smoothly as attention cost grows.
This shows that \rpp{} supports both short and very long contexts without prohibitive overhead.

\subsection{Ablation Studies}
\label{sec:eval-ablation}

\subsubsection{Pipeline Schedule}
\label{sec:eval-ablation-bubble}

\begin{figure}[t]
\centering
\includegraphics[width=\columnwidth]{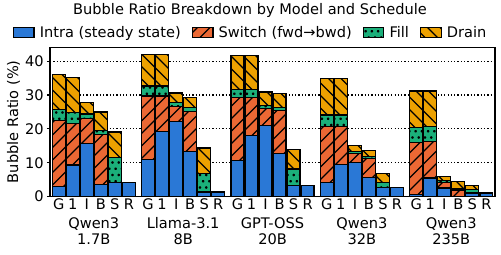}
\par\vspace{-0.1cm}
\caption{\rev{Simulated pipeline bubble ratio of each schedule. G: GPipe, 1: 1F1B, I: Interleaved-1F1B, B: Looped-BFS, S: \rpp{}-sync, R: \rpp{}.}}
\vspace{-0.1cm}
\label{fig:bubble-sim}
\end{figure}

To isolate the contribution of \rpp{}'s pipeline schedule from the end-to-end performance, we simulate the bubble ratio of five pipeline schedules using per-layer timing data collected from the 4090 server.
For each model, we measure forward and backward execution times of each transformer layer and the language model head using a micro-batch size of 4 $\times$ 2048 tokens per micro-batch.
We write a pipeline simulator that faithfully implements each schedule's logic.
We simulate the pipeline execution for 16 micro-batches on 8 GPUs.

\begin{revblock}
Figure~\ref{fig:bubble-sim} shows the simulation result, breaking the bubbles into four categories:
\begin{inparaenum}[(1)]
\item \textit{fill}: filling the pipeline with the first few micro-batches;
\item \textit{switch}: transitioning between the forward and backward passes;
\item \textit{drain}: draining the pipeline for the last few micro-batches; and
\item \textit{intra-stage}: computational imbalance within a single stage.
\end{inparaenum}
Non-overlapped communication contributes less than 1\% of the total runtime across all experiments and is therefore excluded.
\end{revblock}

\rpp{}-sync reduces bubbles by $23\% \sim 55\%$ relative to the best baseline, and its bubble ratio further decreases as deeper models are split into more stages.
\rpp{} with asynchronous optimizer updates virtually eliminates inter-iteration bubbles, driving the absolute bubble ratio down to below $4.5\%$.
These bubbles are all attributed to stage imbalances, which our partitioning algorithm (\S\ref{sec:auto-split}) successfully bounds to a very low level.

We measured the wall-clock time for \rpp{}'s automatic stage-splitting algorithm (\S\ref{sec:auto-split}).
Partitioning Qwen3-1.7B, LLaMA-3.1-8B, GPT-OSS-20B, and Qwen3-32B takes 2.9, 2.9, 2.6, and 5.0 milliseconds, respectively, proving the algorithm is exceedingly fast in practice.
Partitioning the 94 layers Qwen3-235B takes 1.47 seconds, which is still inconsequential compared to the hours required for model training.

\subsubsection{Priority-aware Data Transfer}
\label{sec:eval-ablation-transfer}

\rpp{}'s multi-stream architecture (\S\ref{sec:data-overlap}) overlaps parameter uploads and gradient downloads with GPU computation by interleaving transfers into the idle intervals between activation transfers.
To isolate the contribution of this design, we train Llama3.1-8B on an RTX~4090 and compare three variants:
\begin{inparaenum}[(1)]
\item No Prefetch: transfers are issued sequentially and block computation;
\item Simple Prefetch: transfers begin at the start of each stage without scheduling across micro-batch intervals;
\item \rpp{}: full fine-grained overlap with transfer chunking.
\end{inparaenum}

Figure~\ref{fig:abl-transfer} shows throughput as the microbatch size (mbs) increases from 1 to 10.
\rpp{}'s performance plateaus early at around mbs=3, reflecting that the transfers are already fully hidden at that point.
It delivers $1.34\times$ training throughput compared to the no-prefetch variant and $1.15\times$ throughput compared to the simple prefetch variant.
As mbs grows, each micro-batch provides a longer computation window, allowing the simple prefetch to eventually overlap most of the transfer with computation.
These results confirm that \rpp{}'s data transfer design successfully overlaps data transfer with computations and effectively addresses the head-of-line blocking during data prefetches.

\subsubsection{\lmcp{}}
\label{sec:eval-ablation-optim}

\rpp{}'s \lmcp{} (\S\ref{sec:consistency-protocol}) avoids blocking the main thread during the FP32$\to$FP16 weight copy and gradient collection that bridges the asynchronous optimizer with the GPU workers.
To quantify the benefit, we compare it with the blocking copy baseline that inserts a global barrier before performing the FP32$\to$FP16 weight cast on the main thread (\S\ref{sec:consistency-protocol}).
We measure the per-iteration training time of both methods to quantify the performance difference.

Figure~\ref{fig:copy-overhead} presents the results.
The consistency protocol reduces a substantial overhead of $2.6\sim 14$ seconds per iteration, growing roughly with the size of trainable parameters.
Qwen3-235B-LoRA benefits less because LoRA updates only a small subset of weights and therefore copies less data.
These results confirm that the \lmcp{} is essential for realizing the full benefit of asynchronous optimizer updates.
By replacing the iteration barrier with fine-grained, per-layer event signaling, \rpp{} effectively overlaps weight and gradient synchronization with GPU computation, converting what would otherwise be idle pipeline stalls into productive work.

%% file: sec/discussion.tex
\begin{revblock}
\section{Discussion}
\label{sec:discussion}

\paragraph{Multi-node scaling.}
\rpp{} can be extended to multi-node setups via two approaches.
First, \rpp{} can be combined with ZeRO-1 DP~\cite{rajbhandari2020zeromemoryoptimizationstraining}, using \rpp{} for intra-node PP and ZeRO-1 for inter-node DP.
Based on \rpp{}'s decoupled optimizer worker (\S\ref{sec:overview}), the CPU can perform ZeRO-1's gradient reduce-scatter without interfering with GPU computation, thereby overlapping cross-node communication with computation.
Second, \rpp{}'s pipeline schedule can be adapted by interleaving model layers across multiple nodes.
Within each node, \rpp{}'s core principles---the stateless GPU worker model, the \lmcp{}, optimizer placement, and CPU--GPU transfer coordination---remain applicable and beneficial.
Compared to traditional pipeline parallelism, \rpp{} still offers better scheduling flexibility in the multi-node context, reducing pipeline bubbles and improving GPU load balancing.
The additional overhead is the inter-node transmission of hidden states, which involves minimal communication volume and can be hidden behind computation.
\end{revblock}

%% file: sec/related.tex
\section{Related Work}
\label{sec:related}

\subsection{Pipeline Parallel Schedules}

Synchronous pipeline schedules~\cite{huang2019gpipeefficienttraininggiant,jain2020gems,fan2021dapple,li2021chimera,shoeybi2020megatronlmtrainingmultibillionparameter} are subject to the pipeline bubble problem.
Asynchronous approaches~\cite{harlap2018pipedreamfastefficientpipeline,narayanan2021memoryefficientpipelineparalleldnntraining,chen2018efficient,guan2019xpipe,yang2022group,chen2023elastic,yang2021pipemare} reduce bubbles via weight stashing, and the Zero Bubble Schedule~\cite{qi2023zerobubblepipelineparallelism} reduces bubbles via delaying gradient calculations for weights, but both techniques trade memory consumption for efficiency.
Looped pipeline methods~\cite{lamypoirier2023breadthfirstpipelineparallelism,narayanan2021efficientlargescalelanguagemodel} increase stage counts to reduce bubbles, but they require the stage count to be a multiple of the GPU count, making balanced partitioning increasingly difficult.
\rpp{} takes a different approach by utilizing heterogeneous memory to decouple stages from GPUs, providing a synchronous schedule with flexible stage partitions, and an asynchronous schedule with no GPU memory overhead.

\subsection{Offloading Training Frameworks}

Heterogeneous memory offloading has been widely explored to enable training models that exceed GPU memory.
One line of work offloads model weights and optimizer states to CPU or NVMe~\cite{rajbhandari2021zeroinfinitybreakinggpumemory,stronghold,fang2022parallel,yang2026efficientheterogeneouscodesignfinetuning}.
Another line targets activation offloading, swapping intermediate activations to the host to reduce peak GPU memory~\cite{wang2018superneurons,rhu2016vdnn,zong2023str,bae2021flashneuron}.
More recent systems manage data at tensor granularity to achieve better transfer--compute overlaps~\cite{zhang2023g10,liao2024lohanlowcosthighperformanceframework,ren2021sentinel}.
However, these approaches are predominantly designed for single-GPU or data-parallel settings; scaling them to multiple GPUs incurs substantial communication overhead~\cite{feng2023mobius}.
\rpp{} co-designs distributed training with host-memory offloading, achieving scaling and offloading with negligible overhead.

\begin{revblock}
\subsection{Novelty Relative to Prior Systems}
\label{sec:related-binding}

The primary novelty of \rpp{} lies in \textit{unbinding pipeline stages from specific GPUs} and enabling \textit{flexible pipeline partitioning}.
All prior pipeline training systems bind stages to GPUs to minimize communication overhead, a design choice prevalent in non-offloading systems.
Existing works have overlooked the key insight of \rpp{}: fetching model layers from host memory to the GPU remains compute-bound (\S\ref{sec:cop-roofline}), so computational efficiency is not bottlenecked by weight loading.
Driven by this observation, \rpp{} is the first system to propose and realize dynamic stage dispatching for pipeline-parallel training, and it exploits the resulting scheduling freedom with flexible stage partitioning to reduce pipeline bubbles and improve GPU utilization.
Importantly, efficiently implementing stage dispatching requires addressing the challenges detailed in \S\ref{sec:design}; naively adapting stage dispatching into existing frameworks would yield suboptimal performance.
\end{revblock}